# Temperature dependence of magnetic resonance in ferrimagnetic GdFeCo alloys

Takaya Okuno[1], Se Kwon Kim[2,3,†], Takahiro Moriyama[1], Duck-Ho Kim[1], Hayato Mizuno[1,4], Tetsuya Ikebuchi[1], Yuushou Hirata[1], Hiroki Yoshikawa[5], Arata Tsukamoto[5], Kab-Jin Kim[6], Yoichi Shiota[1], Kyung-Jin Lee[7,8], Teruo Ono[1,9,†]

[1]*Institute for Chemical Research, Kyoto University, Uji, Kyoto 611-0011, Japan*
[2]*Department of Physics and Astronomy, University of California Los Angeles, California 90095, USA*
[3]*Department of Physics and Astronomy, University of Missouri, Columbia, Missouri 65211, USA*
[4]*Institute for Solid State Physics, University of Tokyo, Kashiwa 277-8581, Japan*
[5]*College of Science and Technology, Nihon University, Funabashi, Chiba 274-8501, Japan*
[6]*Department of Physics, Korea Advanced Institute of Science and Technology, Daejeon 34141, Republic of Korea*
[7]*Department of Materials Science & Engineering, Korea University, Seoul 02841, Republic of Korea*
[8]*KU-KIST Graduate School of Converging Science and Technology, Korea University, Seoul 02841, Republic of Korea*
[9]*Center for Spintronics Research Network (CSRN), Graduate School of Engineering Science, Osaka University, Osaka 560-8531, Japan*

†E-mail: kimsek@missouri.edu, ono@scl.kyoto-u.ac.jp

We provide a macroscopic theory and experimental results for magnetic resonances of antiferromagnetically-coupled ferrimagnets. Our theory, which interpolates the dynamics of antiferromagnets and ferromagnets smoothly, can describe ferrimagnetic resonances across the angular momentum compensation point. We also present experimental results for spin-torque induced ferrimagnetic resonance at several temperatures. The spectral analysis based on our theory reveals that the Gilbert damping parameter, which has been considered to be strongly temperature dependent, is insensitive to temperature. We envision that our work will facilitate further investigation of ferrimagnetic dynamics by providing a theoretical framework suitable for a broad range of temperatures.

Antiferromagnets have been gaining much attention in spintronics because of their potential utility for high-speed ultra-dense spintronic devices.[1-4] Due to the antiparallel alignment of adjacent spins, their dynamics is different from that of ferromagnets.[5] One emerging material platform for studying antiferromagnetic dynamics is antiferromagnetically-coupled ferrimagnets,[6-11] for which we can use conventional techniques for ferromagnets owing to small but finite magnetizations. Indeed, recent experiments in such ferrimagnets have found that both field-driven and current-driven domain-wall dynamics are fastest at the angular momentum compensation point $T_A$ where the magnetic dynamics are antiferromagnetic.[12-15] However, the magnetic resonance phenomenon of ferrimagnets (FiMR) has not been fully clarified so far because of insufficient experimental investigations. In the literature, Stanciu *et al.* have studied the laser-induced precession and its decay to equilibrium in ferrimagnets and concluded that the effective Gilbert damping parameter $\alpha$, which governs the dissipation rate of angular momentum, is strongly temperature dependent and increases significantly at $T_A$.[6] However, some recent studies have provided a new perspective on $\alpha$ of ferrimagnets by fully considering the antiferromagnetic dynamics in ferrimagnets. Kamra *et al.* have theoretically revealed that the temperature dependence of FiMR occurs because of the temperature dependence of magnetic dynamics, not because of temperature dependence of $\alpha$.[16] Kim *et al.* have also reported the temperature-insensitive $\alpha$ of ferrimagnets through the DW motion experiments.[17] In this paper, we provide an additional evidence of temperature-insensitive $\alpha$ of ferrimagnets by performing the FiMR experiment analyzed by our macroscopic FiMR theory.

First, we derive the equations for FiMR in a ferrimagnet consisting of two antiferromagnetically-coupled sublattices. Throughout the manuscript, we will focus on the regime where the ferrimagnet is away from the magnetization compensation temperature $T_M$ so that the magnetization is finite and well defined. Our experiments are also performed well within the considered regime as detailed below. To this end, we expand the Landau-Lifshitz-Gilbert-like (LLG-like) equation for ferrimagnet films[12,18-20] at the uniform ground state along the positive in-plane z direction to linear order in the small fluctuations $|n_x|, |n_y| \ll 1$, where the unit vector $\boldsymbol{n}$ represents the Néel order parameter. The resultant equations are given by

$$s_{\text{net}}\dot{n}_x - \alpha s_{\text{total}}\dot{n}_y - \rho \ddot{n}_y = MH_{\text{ext}}n_y, \quad (1)$$

$$s_{\text{net}}\dot{n}_y + \alpha s_{\text{total}}\dot{n}_x + \rho \ddot{n}_x = -M(H_{\text{ext}} + H_{\text{ani}})n_x, \quad (2)$$

where $s_{\text{net}} = s_1 - s_2$ is the net spin density of two sublattices $s_1 > 0$ and $s_2 > 0$, $\alpha$ is the Gilbert damping parameter, $s_{\text{total}} = s_1 + s_2$ is the sum of the magnitudes of the two spin densities, $\rho > 0$ is the moment of inertia for the dynamics (which is inversely proportional to the microscopic exchange field between the two sublattices and describes the antiferromagnetic dynamics of the magnet),[2] $H_{\text{ext}}$ is the external field along the z direction, $H_{\text{ani}}$ is the effective anisotropy field along the x direction perpendicular to the film (including the effect of the demagnetizing field), and $M$ is the magnetization. Here, we are neglecting the terms that are quadratic or higher order in $H_{\text{ext}}$ and the time derivative of the order parameter. The damping term is added by considering the Rayleigh dissipation function $R = \alpha s_{\text{total}} \int dV \, \dot{\mathbf{n}}^2/2$, which is the half of the energy dissipation rate through the magnetic dynamics.[20] Note that the Rayleigh function is defined in terms of $s_{\text{total}}$, not in terms of $s_{\text{net}}$, so that it is well defined even in the vicinity of $T_A$ where $s_{\text{net}}$ vanishes.[17]

To the zeroth-order in the damping parameter $\alpha$, the resonance frequencies for monochromatic solutions to the above equations are given by

$$f_{\pm}^2 = \frac{s_{\text{net}}^2 + \rho M(2H_{\text{ext}} + H_{\text{ani}}) \pm \sqrt{s_{\text{net}}^4 + 2\rho M(s_{\text{net}})^2(2H_{\text{ext}} + H_{\text{ani}}) + \rho^2 M^2 H_{\text{ani}}^2}}{8\pi^2 \rho^2}, \quad (3)$$

where $f_+$ and $f_-$ are the frequencies for higher and lower resonance frequencies for the given field. Far away from $T_A$, where the net spin density $|s_{\text{net}}|$ is sufficiently large, Eq. (1) and the corresponding dynamics are dominated by the first-order time derivative term and thus we can neglect the second-order term by setting $\rho = 0$. In that ferromagnetic limit, the expression for the lower frequency is reduced to that for the ferromagnet resonance frequency:[21]

$$f_{\text{FiM}} = \frac{M}{2\pi|s_{\text{net}}|}\sqrt{H_{\text{ext}}(H_{\text{ext}} + H_{\text{ani}})}. \quad (4)$$

Note that $M/|s_{\text{net}}|$ is the effective gyromagnetic ratio $\gamma_{\text{eff}}$ of the ferrimagnets. As the temperature approaches $T_A$, the net spin density $|s_{\text{net}}|$ decreases and thus the resonance frequency is expected to increase. However, this formula cannot be used in the vicinity of $T_\text{A}$, where $s_\text{net}$ vanishes and thus the second-order term cannot be neglected. Exactly at $T_\text{A}$, the net spin density vanishes $s_\text{net} = 0$, which reduces the obtained resonance frequencies [Eq. (3)] to

$$f_+ = \frac{1}{2\pi}\sqrt{\frac{M(H_\text{ext} + H_\text{ani})}{\rho}}, \quad f_- = \frac{1}{2\pi}\sqrt{\frac{MH_\text{ext}}{\rho}}. \quad (5)$$

Inclusion of the second-order time derivative term $\propto \rho$ in the LLG-like equations [Eq. (1) and Eq. (2)] is necessary to obtain finite resonance frequencies at $T_\text{A}$; otherwise, the LLG-like equations lack in the reactive dynamic term $\propto s_\text{net}$ at $T_\text{A}$ and become unable to describe the ferrimagnetic dynamics properly therein.

Since our experimental results, which are presented below, are performed away from $T_\text{A}$, let us derive the resonance linewidth for ferrimagnets in the ferromagnetic regime. When we include the Gilbert damping term, the resultant linewidth of ferrimagnets $\Delta H$ (half-width-half-maximum) is given by

$$\Delta H \approx \frac{2\pi\alpha}{\gamma_\text{eff}} \frac{s_\text{total}}{|s_\text{net}|} f_\text{FiM}. \quad (6)$$

Therefore $\alpha$ in ferrimagnet is given by

$$\alpha_\text{FiM} \approx \left(\frac{\gamma_\text{eff}}{2\pi}\right)\frac{|s_\text{net}|}{s_\text{total}}\left(\frac{\Delta H}{f_\text{FiM}}\right). \quad (7)$$

Note that both $s_\text{total}$ and $s_\text{net}$ appear in the linewidth expression because 1) the energy dissipation rate is proportional to $s_\text{total}$ since two lattices contribute additively and 2) the resonance frequency is inversely proportional to $s_\text{net}$. On the other hand, in conventional expressions for ferromagnetic resonance, the two spin-density parameters are assumed to be identical, $s_\text{total} = s_\text{net}$, and the corresponding expression $\alpha_\text{FM} \approx \left(\frac{\gamma_\text{eff}}{2\pi}\right)\left(\frac{\Delta H}{f_\text{FiM}}\right)$ was used to analyze the magnetic resonance of ferrimagnets in the previous reports.[6,7] Below, these two expressions for the Gilbert damping parameters, $\alpha_\text{FiM}$ and $\alpha_\text{FM}$, will be compared based on our experimental results.

We experimentally investigated the FiMR in the GdFeCo compounds by using the homodyne technique[22-24] as shown in Fig 1. For this study, we used a 5-nm SiN/10-nm $Gd_{25.0}Fe_{65.6}Co_{9.4}$/5-nm Pt/100-nm SiN/Si substrate film. The film was patterned into a 10-µm-wide and 10-µm-long strip pattern structure using optical lithography and Ar ion milling. A coplanar waveguide made of 100-nm Au/5-nm Ti were deposited at the ends of the strip.

The measurements were performed by sweeping an external magnetic field $H_{\text{ext}}$ at a fixed rf current $I_{\text{rf}}$ (frequency $f = 4 - 18$ GHz). $H_{\text{ext}}$ was applied in-plane 45° away from the long axis of the strip.

Figure 2a shows the FiMR spectra at several temperatures $T$ between 220 K and 295 K. Although a single peak was clearly observed at 295 K, a second peak was also observed at $H_{\text{ext}} \approx 50$ mT when $T$ is lower than 240 K. Note that the spontaneous magnetization lies in the sample plane at $T = 295$ K while it becomes perpendicular to the plane when $T \leq 240$ K. Thus, the two resonance peaks when $T \leq 240$ K originate from the magnetic resonance of perpendicular ($H_{\text{ext}} \approx 50$ mT) and in-plane (higher field) magnetizations, respectively. Here we focus on the resonance peak originating from in-plane magnetization, so we cut off the low-field regime to exclude the resonance peak from perpendicular magnetization and fit those spectra in Fig. 2a by the combination of symmetric and anti-symmetric Lorentzian functions, from which the resonance parameters are obtained.[22,23]

Figures 2b and 2c show the resonance frequency $f_{\text{res}}$ as a function of the resonance field $H_{\text{res}}$ and the spectral linewidth $\Delta H$ (half-width-half-maximum) as a function of $f_{\text{res}}$, respectively. Firstly, we analyze these data using the conventional expressions of ferromagnetic resonance,[21,25]

$$f_{\text{res}} = \frac{g_{\text{eff}}\mu_B}{h}\sqrt{H_{\text{res}}(H_{\text{res}} + H_{\text{ani}})}, \quad (8)$$

$$\Delta H = \frac{\alpha_{\text{FM}}}{(g_{\text{eff}}\mu_B/h)}f_{\text{res}} + \Delta H_0. \quad (9)$$

Here, $g_{\text{eff}}$ is the effective Landé g-factor, $\mu_B$ is the Bohr magneton, $h$ is the Planck's constant, $H_{\text{ani}}$ is the effective anisotropy field including the demagnetization field, $\alpha_{\text{FM}}$ is the effective Gilbert damping parameter defined as in Ref. 6, and $\Delta H_0$ is a frequency-independent linewidth known as the inhomogeneous broadening, which originates from magnetic non-uniformity.[25] Equation (8) can be matched with Eq. (4) once we identify $g_{\text{eff}}\mu_B/\hbar$ as the effective gyromagnetic ratio $M/|s_{\text{net}}|$ ($\hbar = h/2\pi$ is the reduced Planck's constant) and $H_{\text{res}}$ as $H_{\text{ext}}$. The $H_{\text{res}}$ vs $f_{\text{res}}$ shown in Fig. 2b are well fitted by Eq. (8), indicated by the solid lines, and $g_{\text{eff}}$ and $H_{\text{ani}}$ are obtained as the fitting parameters. Figures 3a and 3b show $g_{\text{eff}}$ and $H_{\text{ani}}$ as a function of $T$, respectively. It is found that $g_{\text{eff}}$ remarkably increases as $T$ decreases. Since the $T_A$ of the device is estimated to be 160 K (see below the estimation method), the result shows that $g_{\text{eff}}$ increases as $T$ approaches $T_A$. Note that the drastic decrease in $H_{\text{ani}}$ with decreasing $T$ (Fig. 3b) is attributed to the change in magnetic anisotropy from in-plane (295 K) to perpendicular (220 K) direction as mentioned above. The $f_{\text{res}}$ vs $\Delta H$ shown in Fig. 2c are well fitted by Eq. (9), indicated by the solid lines, and $\alpha_{\text{FM}}$ and $\Delta H_0$ are obtained as the fitting parameters. Figures 3c and 3d show $\alpha_{\text{FM}}$ and $\Delta H_0$ as a function of $T$, respectively. It is found that $\alpha_{\text{FM}}$ increases significantly as $T$ decreases, i.e. as $T$ approaches $T_A$. The $T$ dependences of $g_{\text{eff}}$ and $\alpha_{\text{FM}}$ are in good agreement with the previous papers.[6,7,27] According to the previous papers,[6,7] the $T$ dependences of $g_{\text{eff}}$ and $\alpha_{\text{FM}}$ are understood in terms of that of the net angular momentum $s_{\text{net}}$; both $g_{\text{eff}}\mu_B/\hbar = M_{\text{net}}/s_{\text{net}}$ and $\alpha_{\text{FM}}$ [from Eq. (9)] are ill-defined at $T_A$ where $s_{\text{net}}$ vanishes, which makes the theory based on ferromagnets invalid therein. However, as shown in the discussion of our theory for FiMR, by defining the Gilbert damping parameter in the Rayleigh dissipation function $R = \alpha s_{\text{total}} \int dV \, \dot{\mathbf{n}}^2/2$ in such a way that the damping parameter is always well-defined, the resonance frequency and the linewidth of FiMR can be described properly across $T_A$. In order to test whether our theory can explain the experimental results, we analyze those data in Figs. 2b and 2c based on our theory.

As mentioned in the theory part, the ferrimagnetic resonance frequency in the ferromagnetic limit is reduced to the conventional ferromagnetic case, while the spectral linewidth is modified by including the additional term $s_{\text{net}}/s_{\text{total}}$. Therefore, the Gilbert damping parameter [Eq. (7)] in our theory [Eq. (1) and Eq. (2)] for the dynamics of ferrimagnets can be obtained by the following expression:

$$\alpha_{\text{FiM}} = \alpha_{\text{FM}} \left| \frac{s_{\text{net}}}{s_{\text{total}}} \right|. \quad (10)$$

To obtain $\alpha_{\text{FiM}}$ from Eq. (7) and Fig. 2c, $s_{\text{net}}/s_{\text{total}}$ needs to be acquired. Although the net spin density $s_{\text{net}}$ is easy to obtain from the effective gyromagnetic ratio, the total spin density $s_{\text{total}}$ is not straightforward to obtain. To solve this problem, we perform the following analysis. The effective net gyromagnetic ratio satisfies the following equation;[6,7]

$$\frac{g_{\text{eff}}\mu_B}{\hbar} = \frac{M_{\text{net}}}{s_{\text{net}}} = \frac{M_{\text{FeCo}} - M_{\text{Gd}}}{\dfrac{M_{\text{FeCo}}}{(g_{\text{FeCo}}\mu_B/\hbar)} - \dfrac{M_{\text{Gd}}}{(g_{\text{Gd}}\mu_B/\hbar)}}. \quad (11)$$

Here, $M_{\text{FeCo}}$ ($M_{\text{Gd}}$) is the magnetizations of transition metal (rare-earth metal), and $g_{\text{FeCo}}$ ($g_{\text{Gd}}$) is the Landé $g$-factor of transition metal (rare-earth metal) sublattice. $g_{\text{eff}}$ is shown in Fig. 3a, $g_{FeCo}$ and $g_{Gd}$ are obtained from literature ($g_{FeCo} \sim 2.2$ and $g_{Gd} \sim 2.0$).[28-30] Two quantities can be measured directly: $M_{net}$ is independently measured by SQUID as shown in Fig. 4a and $s_{net}$ can be obtained from the effective gyromagnetic ratio when the ferrimagnet is well within the ferromagnetic regime. With the measured values of $M_{net} = M_{FeCo} - M_{Gd}$ and $s_{net} = \frac{M_{FeCo}}{(g_{FeCo}\mu_B/\hbar)} - \frac{M_{Gd}}{(g_{Gd}\mu_B/\hbar)}$, we can obtain the magnetizations of two sublattices, $M_{FeCo}$ and $M_{Gd}$, and also the spin densities of two sublattices, $s_{FeCo}$ and $s_{Gd}$. From these results, we can obtain the total spin density $s_{total} = s_{FeCo} + s_{Gd}$. Figures 4a and 4b show $M_{FeCo}$ and $M_{Gd}$, and $s_{net} = s_{FeCo} - s_{Gd}$ and $s_{total} = s_{FeCo} + s_{Gd}$ as a function of $T$, respectively. Note that the $T_M$ (110 K) determined by SQUID (Fig. 4a) and the $T_A$ (160 K) roughly estimated from the $T$ dependence of $s_{net}$ (Fig. 4b) are clearly different,[12,31] which supports the validity of this analysis. Finally, by substituting $s_{net}$ and $s_{total}$ into Eq. (10), the damping parameter $\alpha_{FiM}$ is obtained as shown in Fig. 4c. It can be clearly seen that $\alpha_{FiM}(\approx 0.01)$ is insensitive to $T$, in sharp contrast to $\alpha_{FM}$ which significantly increases as $T$ approaches $T_A$. Note that Eq. (10) is valid only in the ferromagnetic limit and, therefore, it is necessary to confirm that the measured temperature range (220 – 295 K) is in the deep ferromagnetic regime. This would be guaranteed by the facts that 1) Fig. 4b shows $T_A \sim 160$ K, which is far below the lowest $T$ in our measurements (220 K), and 2) the resonance frequency at $T_A$ is expected to be similar to or larger than about 50 GHz under 300 mT,[6] which is much larger than the experimentally obtained resonance frequency at 220 K (12 GHz under 300 mT).

The observation that $\alpha_{FiM}$ is insensitive to $T$ indicates that the $T$ dependence of the spectral linewidth in FiMR is attributed to the $T$ dependence of the net spin density $s_{net}$ instead of that of the effective Gilbert damping parameter. This conclusion is consistent with some recent papers,[16,17] but it is in sharp contrast to the interpretation of the previous reports,[6,7] where the $T$ dependence of the spectral linewidth in FiMR was attributed to the change of the effective Gilbert damping parameter. Our results provide an additional and clear evidence that properly defined Gilbert damping parameter $\alpha_{FiM}$ of ferrimagnets is insensitive to temperature, supporting the validity of these papers.[16,17] Here, we would like to mention that, even though Fig. 4c is an evidence for the temperature-insensitive $\alpha_{FiM}$ of ferrimagnets, it lacks the information of $\alpha_{FiM}$ in the vicinity of $T_A$. However, obtaining $\alpha_{FiM}$ in the vicinity of $T_A$ based on FiMR experiments is challenging because 1) experimental observation of ferrimagnetic resonance at $T_A$ which was measured to be larger than 50 GHz for certain ferrimagnets[6)] is expected to be difficult with homodyne detection technique (40 GHz at maximum in our measurement system) and 2) obtaining the necessary parameter $s_\text{total}$ is difficult because the net spin density $s_\text{net}$ cannot be obtained from the effective gyromagnetic ratio in the vicinity of $T_\text{A}$. Therefore, we believe that Fig. 4c serves as a good experimental evidence to conclude that $\alpha_\text{FiM}$ of ferrimagnets is insensitive to temperature.

In conclusion, we have provided the macroscopic theoretical description of ferrimagnetic resonance and experimental results that support it. Our theory shows that the resonance frequency and the spectral linewidth of ferrimagnetic resonance can be described well across the angular momentum compensation point, by adding the antiferromagnetic-like inertial term to the equations of motion and by defining the Gilbert damping parameter properly through the Rayleigh dissipation function. Moreover, we performed the spin-torque induced ferrimagnetic resonances at various temperatures and successfully observed that the resonance frequency and the linewidth depend on temperature. By analyzing the spectrum based on our theory, we found that the Gilbert damping parameter in ferrimagnets is insensitive to temperature, which has been considered to be strongly temperature-dependent. Our work introduces a new framework for studying ferrimagnetic resonance that allows us to interpret the ferrimagnetic dynamics for a wide range of temperatures.

**Acknowledgments**

This work was supported by the JSPS KAKENHI (Grants No. 15H05702, No. 17H04924, No. 17H05181, No. 26103002, and No. 26103004), Collaborative Research Program of the Institute for Chemical Research, Kyoto University, and R & D project for ICT Key Technology of MEXT from the Japan Society for the Promotion of Science (JSPS). This work was partly supported by The Cooperative Research Project Program of the Research Institute of Electrical Communication, Tohoku University. S. K. K. was supported by the startup fund at the University of Missouri. D. H. K. was supported as an Overseas Researcher under the Postdoctoral Fellowship of JSPS (Grant No. P16314). K. J. L. was supported by the National Research Foundation of Korea (2017R1A2B2006119). K. J. K. was supported by the KAIST-funded Global Singularity Research Program for 2019.

**Figure Captions**

**Fig. 1.** The schematic illustration of the device and the measurement setup. The direction of the external magnetic field $H_{ext}$ and the AC current $I_{rf}$ are indicated. $H_{ext}$ was applied in-plane 45° away from the long axis of the strip.

**Fig. 2.** (a) The ferrimagnetic resonance spectra as a function of the external magnetic field $H_{ext}$ at several temperatures from 220-295 K. The emerging peak at $H_{ext} \approx 50$ mT below 240 K is attributed to the out-of-plane resonance peak and are neglected in this study. (b) The resonance frequency $f_{res}$ as a function of the resonance magnetic field $H_{res}$. The solid lines are the fitting results by Eq. (8). (c) The spectral linewidth $\Delta H$ as a function of $f_{res}$. The solid lines are the fitting results by Eq. (9).

**Fig. 3.** Resonance parameters as a function of temperature extracted by the fitting in Fig. (2). (a) The effective Landé $g$-factor $g_{eff}$. (b) The effective anisotropy field $H_{ani}$. (c) The effective Gilbert damping parameter $\alpha_{FM}$. (d) The frequency-independent linewidth $\Delta H_0$.

**Fig. 4.** (a) The net magnetization $M_{net}$ and the magnetizations of two sublattices $M_{FeCo}$ and $M_{Gd}$ as functions of temperature. (b) The net spin density $s_{net}$, the spin densities of two sublattices $s_{FeCo}$ and $s_{Gd}$, and the sum of the magnitudes of the two spin densities $s_{total}$ as functions of temperature. (c) The effective Gilbert damping parameter $\alpha_{FM}$ and the properly defined Gilbert damping parameter of ferrimagnets $\alpha_{FiM}$ as functions of temperature.

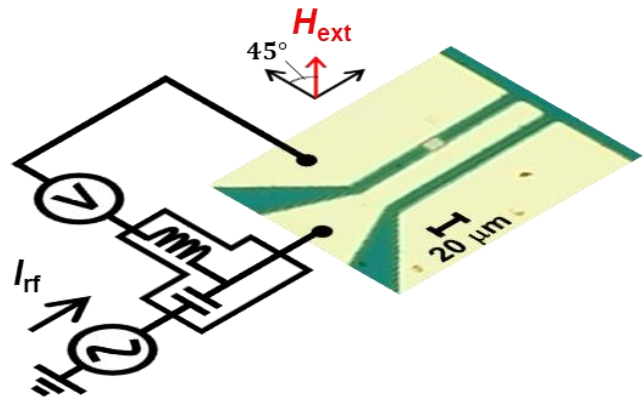

Fig. 1.

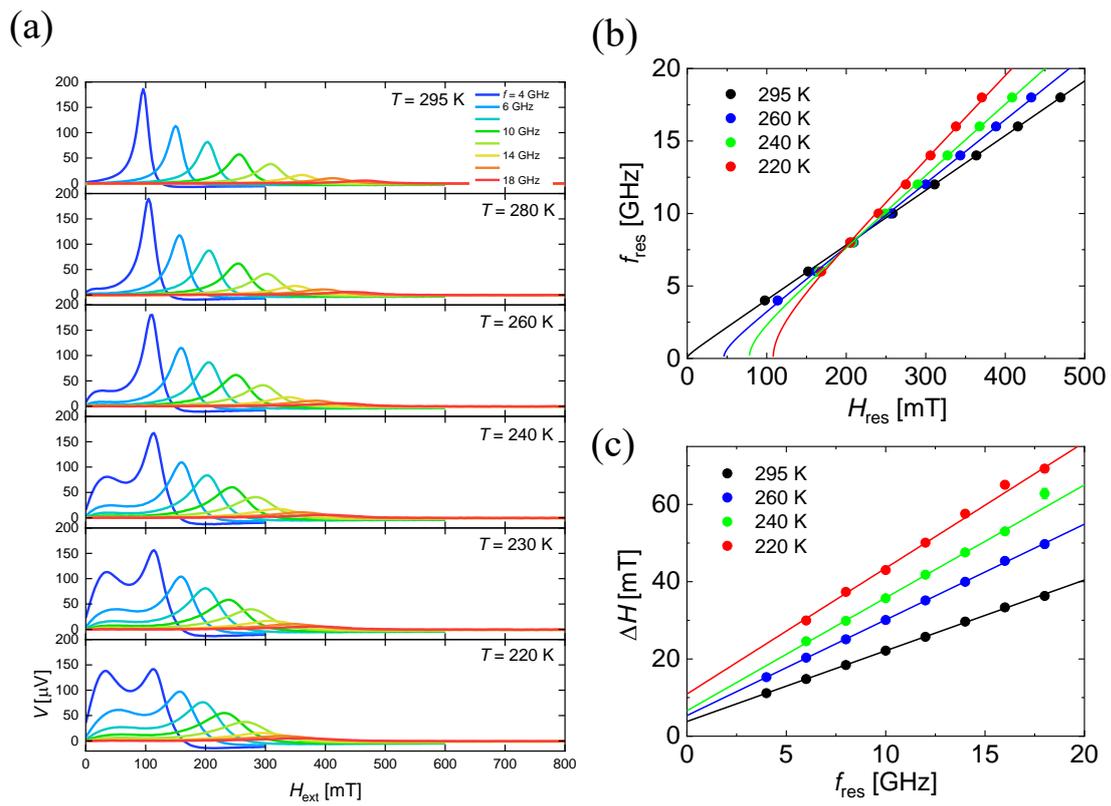

Fig. 2.

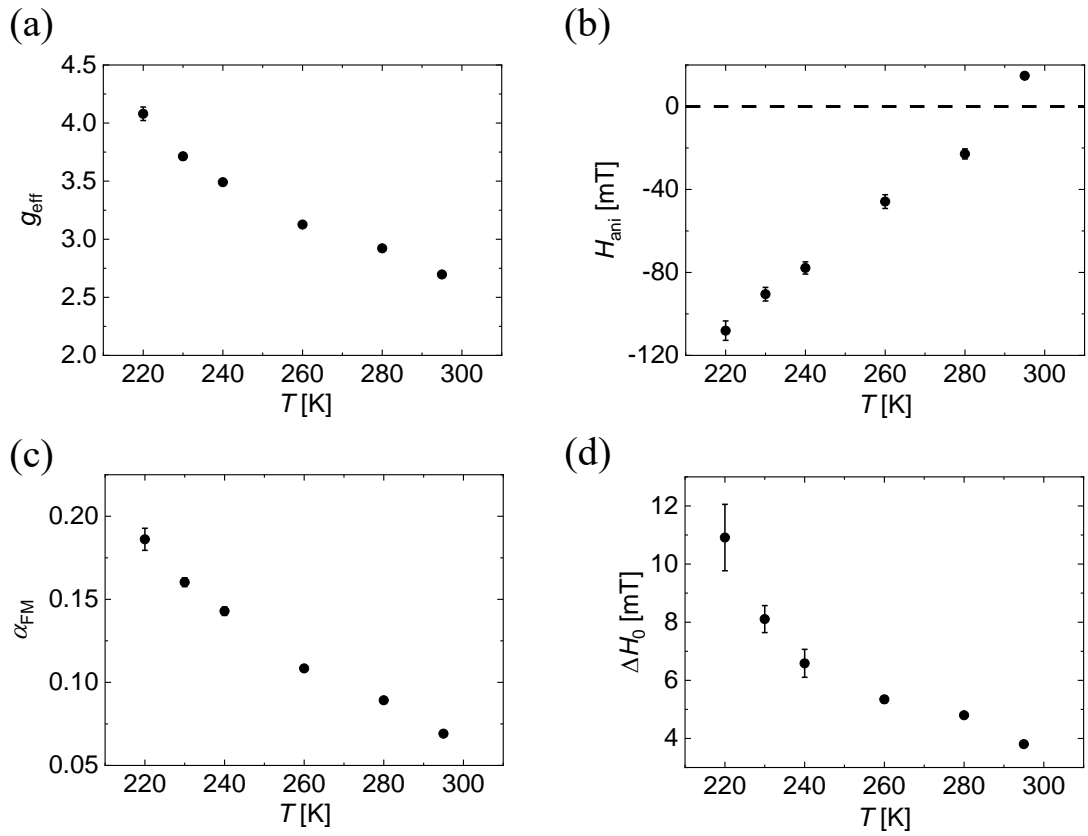

Fig. 3.

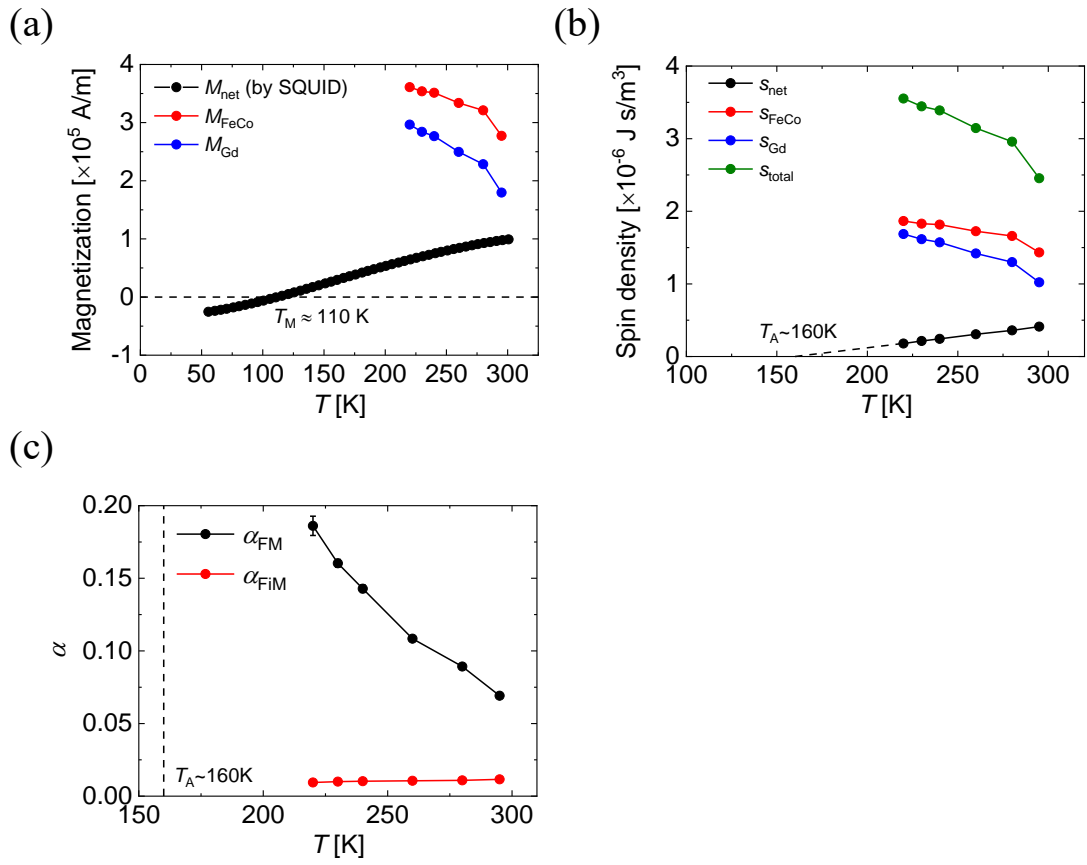

Fig. 4.